\documentclass[twocolumn,tighten]{aastex63}
\usepackage{color}
\usepackage{multirow}

\shortauthors{Sano, Inoue, Tokuda et al.}
\shorttitle{ALMA CO Observations of the Gamma-Ray Supernova Remnant RX~J1713.7$-$3946}

\received{2020 October 7}
\revised{2020 November 3}
\accepted{2020 November 5}

\begin{document}
\title{ALMA CO Observations of the Gamma-Ray Supernova Remnant RX~J1713.7$-$3946:\\Discovery of Shocked Molecular Cloudlets and Filaments at 0.01~pc scales}


\author[0000-0003-2062-5692]{H. Sano}
\affiliation{National Astronomical Observatory of Japan, Mitaka, Tokyo 181-8588, Japan; hidetoshi.sano@nao.ac.jp}
\affiliation{Department of Physics, Nagoya University, Furo-cho, Chikusa-ku, Nagoya 464-8601, Japan}
\affiliation{Institute for Advanced Research, Nagoya University, Furo-cho, Chikusa-ku, Nagoya 464-8601, Japan}

\author{T. Inoue}
\affiliation{Department of Physics, Nagoya University, Furo-cho, Chikusa-ku, Nagoya 464-8601, Japan}

\author[0000-0002-2062-1600]{K. Tokuda}
\affiliation{National Astronomical Observatory of Japan, Mitaka, Tokyo 181-8588, Japan; hidetoshi.sano@nao.ac.jp}
\affiliation{Department of Physical Science, Graduate School of Science, Osaka Prefecture University, 1-1 Gakuen-cho, Naka-ku, Sakai 599-8531, Japan}

\author[0000-0002-4383-0368]{T. Tanaka}
\affiliation{Department of Physics, Kyoto University, Kitashirakawa Oiwake-cho, Sakyo, Kyoto 606-8502, Japan}

\author{R. Yamazaki}
\affiliation{Department of Physics and Mathematics, Aoyama Gakuin University, 5-10-1 Fuchinobe, Sagamihara 252-5258, Japan}
\affiliation{Institute of Laser Engineering, Osaka University, 2-6 Yamadaoka, Suita, Osaka 565-0871, Japan}

\author[0000-0003-4366-6518]{S. Inutsuka}
\affiliation{Department of Physics, Nagoya University, Furo-cho, Chikusa-ku, Nagoya 464-8601, Japan}

\author{F. Aharonian}
\affiliation{Dublin Institute for Advanced Studies, 31 Fitzwilliam Place, Dublin 2, Ireland}
\affiliation{Max-Planck-Institut f\"{u}r Kernphysik, P.O. Box 103980, D 69029 Heidelberg, Germany}
\affiliation{Gran Sasso Science Institute, 7 viale Francesco Crispi, 67100 L’Aquila, Italy}

\author[0000-0002-9516-1581]{G. Rowell}
\affiliation{School of Physical Sciences, The University of Adelaide, North Terrace, Adelaide, SA 5005, Australia}

\author[0000-0002-4990-9288]{M. D. Filipovi{\'c}}
\affiliation{Western Sydney University, Locked Bag 1797, Penrith South DC, NSW 1797, Australia}

\author[0000-0001-8296-7482]{Y. Yamane}
\affiliation{Department of Physics, Nagoya University, Furo-cho, Chikusa-ku, Nagoya 464-8601, Japan}

\author[0000-0002-2458-7876]{S. Yoshiike}
\affiliation{Department of Physics, Nagoya University, Furo-cho, Chikusa-ku, Nagoya 464-8601, Japan}

\author[0000-0003-2762-8378]{N. Maxted}
\affiliation{School of Science, University of New South Wales, Australian Defence Force Academy, Canberra, ACT 2600, Australia}

\author[0000-0003-1518-2188]{H. Uchida}
\affiliation{Department of Physics, Kyoto University, Kitashirakawa Oiwake-cho, Sakyo, Kyoto 606-8502, Japan}

\author[0000-0003-0324-1689]{T. Hayakawa}
\affiliation{Department of Physics, Nagoya University, Furo-cho, Chikusa-ku, Nagoya 464-8601, Japan}

\author[0000-0002-1411-5410]{K. Tachihara}
\affiliation{Department of Physics, Nagoya University, Furo-cho, Chikusa-ku, Nagoya 464-8601, Japan}

\author{Y. Uchiyama}
\affiliation{Department of Physics, Rikkyo University, 3-34-1 Nishi Ikebukuro, Toshima-ku, Tokyo 171-8501, Japan}

\author{Y. Fukui}
\affiliation{Department of Physics, Nagoya University, Furo-cho, Chikusa-ku, Nagoya 464-8601, Japan}
\affiliation{Institute for Advanced Research, Nagoya University, Furo-cho, Chikusa-ku, Nagoya 464-8601, Japan}

\begin{abstract}
RX~J1713.7$-$3946 is a unique core-collapse SNR that emits bright TeV gamma-rays and synchrotron X-rays caused by cosmic rays, in addition to interactions with interstellar gas clouds. We report here on results of ALMA $^{12}$CO($J$=1--0) observations toward the northwestern shell of the SNR. We newly found three molecular complexes consisting of dozens of shocked molecular cloudlets and filaments with typical radii of $\sim$0.03--0.05~pc and densities of $\sim$$10^4$~cm$^{-3}$. These cloudlets and filaments are located not only along synchrotron X-ray filaments, but also in the vicinity of X-ray hotspots with month or year-scale time variations. We argue that X-ray hotspots were generated by shock-cloudlet interactions through magnetic-field amplification up to mG. The ISM density contrast of $\sim$$10^5$, coexistence of molecular cloudlets and low-density diffuse medium of $\sim$0.1~cm$^{-3}$, is consistent with such a magnetic field amplification as well as a wind-bubble scenario. The small-scale cloud structures also affect hadronic gamma-ray spectra considering the magnetic field amplification on surface and inside clouds.

\end{abstract}

\section{Introduction}\label{sec:intro}
Galactic cosmic rays are believed to be accelerated by shockwaves in supernova remnants (SNR) via the diffusive shock acceleration \citep[DSA;][]{1978MNRAS.182..147B,1978ApJ...221L..29B}, and cosmic ray induced non-thermal radiation has been predicted and/or detected from various SNRs \citep[e.g.,][]{1994A&A...285..645A,1994A&A...287..959D}. In general, the classical DSA theory assumes uniform density distribution of the interstellar medium (ISM) surrounding SNRs. However, observational results indicated that non-thermal X-ray and/or gamma-ray bright SNRs are tightly interacting with dense and clumpy gaseous medium such as molecular and atomic clouds \citep[e.g.,][]{2003PASJ...55L..61F,2017ApJ...850...71F,2008A&A...481..401A,2016ApJ...826...34Z,2020ApJ...902...53S}. A key issue at present is how shock-cloud interaction affects radiation processes of the non-thermal radiation, as well as acceleration mechanisms of cosmic rays beyond the DSA.

\begin{figure*}[]
\begin{center}
\includegraphics[width=\linewidth,clip]{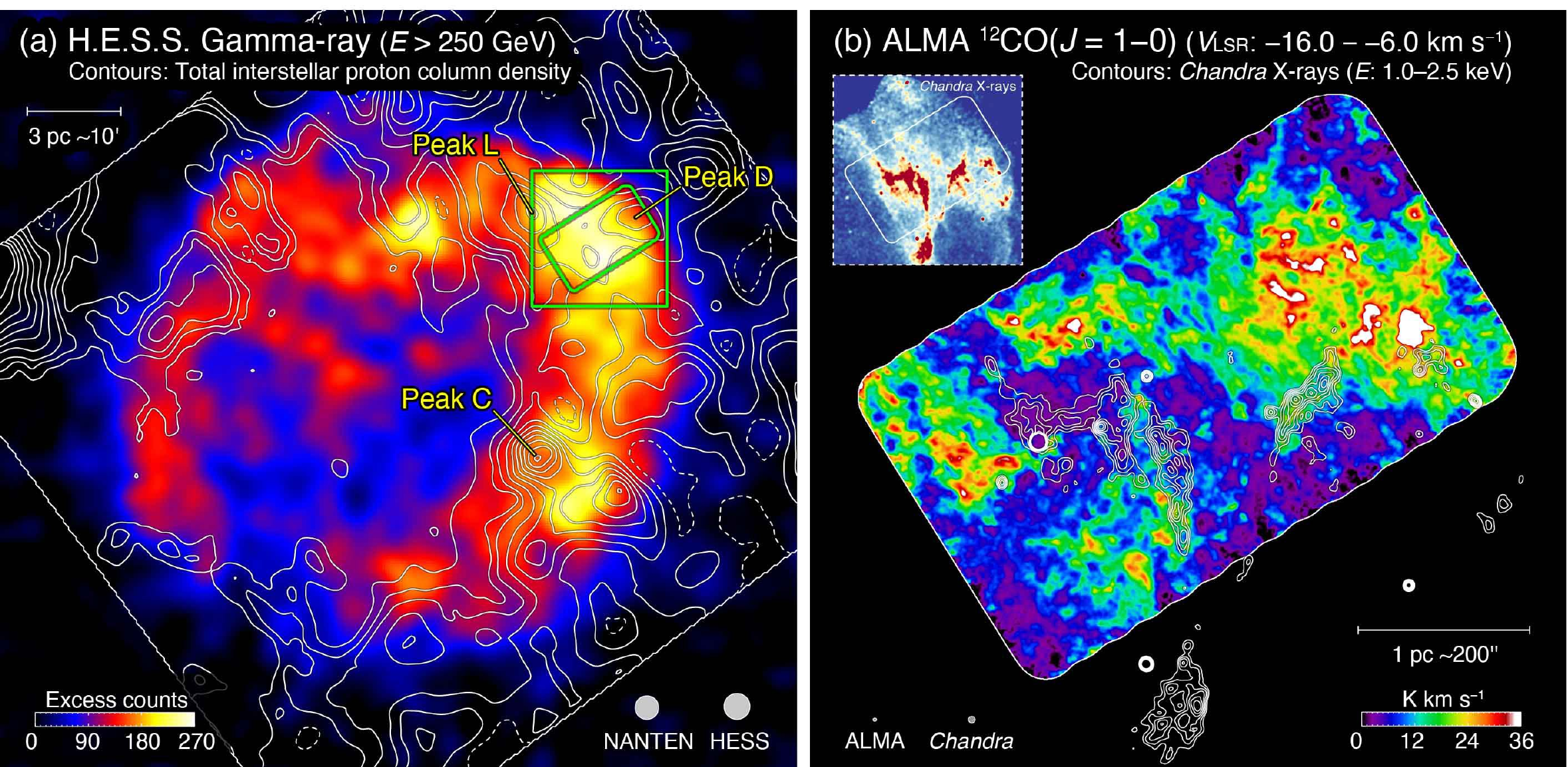}
\caption{(a) TeV gamma-ray excess map of RX~J1713.7$-$3946 in equational coordinates obtained by H.E.S.S. \citep{2018A&A...612A...6H}. Superposed contours indicate the ISM proton column density \citep{2012ApJ...746...82F}. The contour levels are 6, 7, 8, 10, 12, 14, 16, 18, 20, and $22 \times 10^{21}$ cm$^{-2}$. Green rectangles represents the ALMA observed area (small rectangle) and to be presented area in Figure \ref{fig1}b. The positions of CO peaks C, D, and L are shown. (b) ALMA CO map toward northwest of RX~J1713.7$-$3946. Superposed contours indicate the X-ray intensity obtained with {\it{Chandra}} \citep{2007Natur.449..576U}. The lowest contour and contour intervals are $9.5 \times 10^{-8}$ and $9.5 \times 10^{-9}$~photons~cm$^{-2}$~s$^{-1}$~pixel$^{-1}$, respectively. The colored image in the top left indicates {\it{Chandra}} X-ray map in the same region as shown in Figure \ref{fig1}b.}
\label{fig1}
\end{center}
\end{figure*}%

The core-collapse SNR RX~J1713.7$-$3946 (a.k.a. G347.3$-$0.5) provides the best laboratory to test the effect because of its bright non-thermal X-rays and TeV gamma-rays \citep[e.g.,][]{1997PASJ...49L...7K,2004Natur.432...75A}, in addition to certain interactions with dense gas clouds at the closed distance of 1~kpc \citep[e.g.,][]{2003PASJ...55L..61F,2005ApJ...631..947M,2010ApJ...724...59S,2012MNRAS.422.2230M} and its young age of $\sim$1600~yr \citep[][]{1997A&A...318L..59W,2003PASJ...55L..61F,2016PASJ...68..108T}. The X-rays are dominated by synchrotron radiation up to 120~keV \citep[e.g.,][]{1999ApJ...525..357S,2008ApJ...685..988T,2019ApJ...877...96T,2019MNRAS.489.1828K}. {\it{Chandra}} observations discovered year-scale time variability of X-ray hotspots on the order of 10~arcsec or 0.05~pc widths as well as X-ray filaments of $\sim$0.1--0.2~pc widths, indicating efficient cosmic-ray acceleration with amplified magnetic field \citep{2007Natur.449..576U,2020ApJ...899..102H}. The TeV gamma-ray observations of RX~J1713.7$-$3946 revealed its shell-like morphology with photon energies from 200~GeV to 40~TeV \citep{2004Natur.432...75A,2006A&A...449..223A,2007A&A...464..235A,2018A&A...612A...6H}. Despite a number of efforts to model the broad-band spectra, the origin of gamma-rays---hadronic, leptonic, or a combination of both---was not clearly established because all scenarios could reproduce the observed spectra (see \citeauthor{2018A&A...612A...6H} \citeyear{2018A&A...612A...6H} and references therein).

\begin{figure*}[]
\begin{center}
\includegraphics[width=\linewidth,clip]{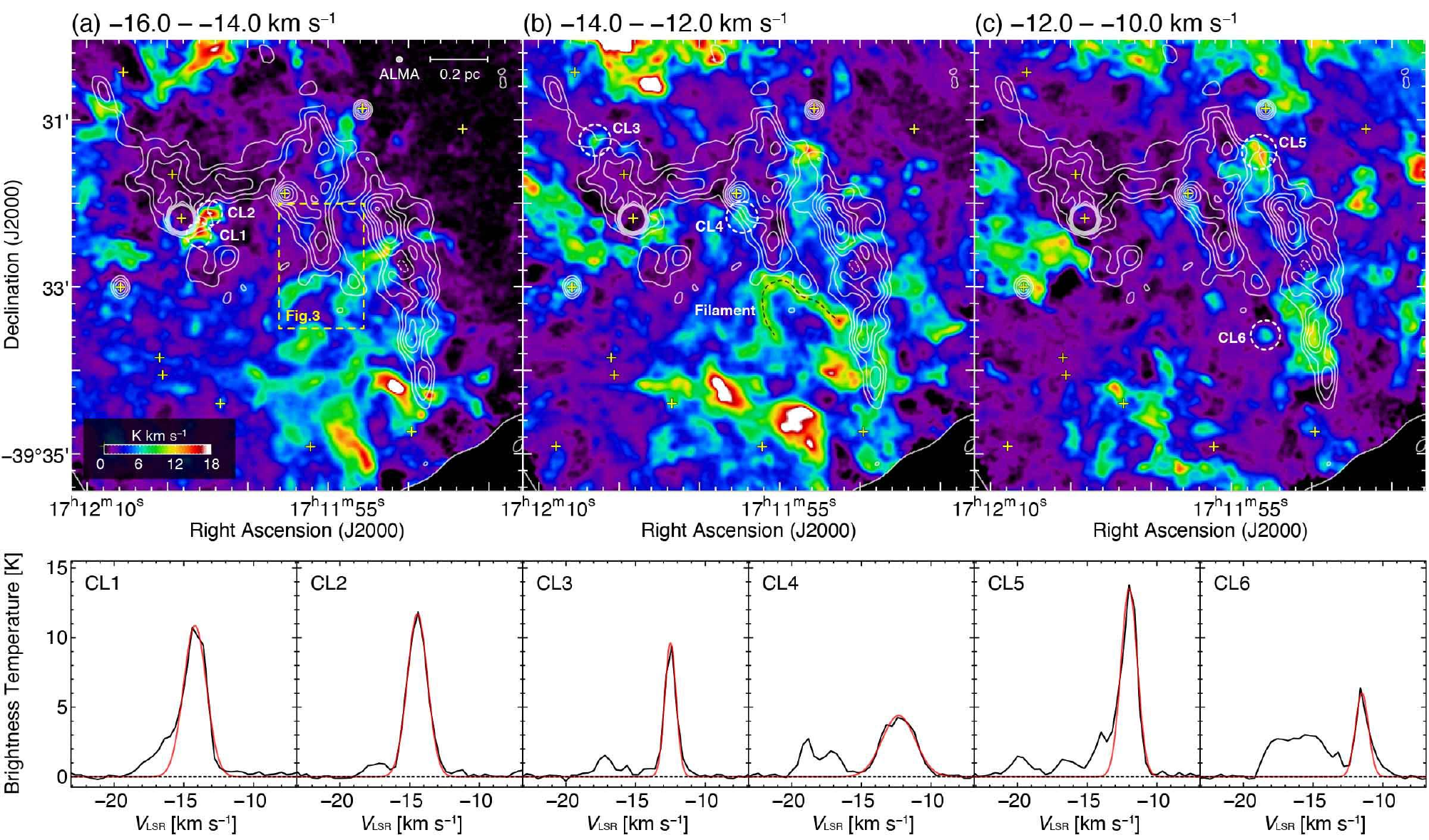}
\caption{({\it{top panels}}) ALMA CO velocity channel maps toward {RXJ1713NW-East}. Each panel shows CO intensity distribution integrated every 2.0~km~s$^{-1}$ in the velocity range from $-16.0$ to $-10.0$~km~s$^{-1}$. Superposed contours indicate the same X-ray intensity as shown in Figure \ref{fig1}b. Yellow crosses indicate positions of X-ray hot spots identified by \cite{2020ApJ...899..102H}. The typical CO cloudlets, CL1--6, and a CO filament are also indicated. ({\it{bottom panels}}) ALMA CO profiles of CL1--6 (black solid lines). Each spectrum is fitted by a single Gaussian kernel using the least-squares method.}
\label{fig2}
\end{center}
\end{figure*}%

Investigating the interstellar gas associated with RX~J1713.7$-$3946 holds a key to understanding the radiation processes and efficient acceleration of cosmic-rays. \cite{2012ApJ...746...82F} presented a good spatial correspondence between the TeV gamma-rays and ISM proton column density. This provides an essential condition for gamma-rays to be predominantly of hadronic origin. \cite{2010ApJ...724...59S,2013ApJ...778...59S} found limb-brightening of synchrotron X-rays toward shocked molecular clouds with a size of $\sim$3~pc, where the small photon indices of synchrotron X-rays are seen \citep{2015ApJ...799..175S,2018PASJ...70...77O}. The authors proposed a possible scenario that the shock-cloud interaction generates turbulence that enhances magnetic field and synchrotron X-rays on the surface of  shocked clouds. Moreover, the enhanced turbulence and/or magnetic field may re-accelerate electrons into higher energy. This interpretation is also supported by numerical results, and well explains the observed broad-band spectra without bright thermal X-ray{s} \citep[e.g.,][]{2009ApJ...695..825I,2012ApJ...744...71I,2019MNRAS.487.3199C}. However, previous studies could not spatially resolve tiny molecular clouds relating to the X-ray hotspots {and filaments} at 0.05~pc scales which were predicted by \cite{2009ApJ...695..825I}.

We report here on results of ALMA CO observations toward the northwestern shell of RX~J1713.7$-$3946 with a spatial resolution of $\sim$0.02~pc. Our findings for complexes of molecular cloudlets at 0.01~pc provide a new perspective on the ISM surrounding core-collapse SNRs.

\section{Results}\label{sec:results}
Figure \ref{fig1}a shows the overall TeV gamma-ray morphology of RX~J1713.7$-$3946 \citep[][]{2018A&A...612A...6H}. The TeV gamma-ray shell shows a good spatial correspondence with the total interstellar proton column density contours \citep{2012ApJ...746...82F}. For the dense star-forming core named as peak C, TeV gamma-ray intensity is reduced respect to its surroundings. Note that the brightest TeV gamma-ray spot is located in the northwestern shell, especially in the intermediate region of peaks D and L. The intercloud region also bright in X-rays with filamentary structures.

\begin{deluxetable*}{lcccccccc}[]
\tablecaption{Physical properties of typical CO cloudlets associated with RX~J1713.7$-$3946}
\tablehead{\multicolumn{1}{c}{Name} & $\alpha_{\mathrm{J2000}}$ & $\delta_{\mathrm{J2000}}$ & $T_{\mathrm{mb}} $ & $V_{\mathrm{peak}}$ & $\Delta V$ & Size & Mass & $n(\mathrm{H_2})$\\
& ($^{\mathrm{h}}$ $^{\mathrm{m}}$ $^{\mathrm{s}}$) & ($^{\circ}$ $\arcmin$ $\arcsec$) & (K) & \scalebox{0.9}[1]{(km $\mathrm{s^{-1}}$)} & \scalebox{0.9}[1]{(km $\mathrm{s^{-1}}$)} & (pc) & ($M_\sun $) & (cm$^{-3}$)\\
\multicolumn{1}{c}{(1)} & (2) & (3) & (4) & (5) & (6) & (7) & (8) & (9)}
\startdata
CL1 ............ & 17 12 02.95 & $-$39 32 20.4 & $10.9 \pm 0.6$ & $-14.20 \pm 0.05$ & $1.94 \pm 0.14$ & 0.08 & 0.37 & $2.6\times10^4$\\
CL2 ............ & 17 12 02.50 & $-$39 32 06.7 &$11.7 \pm 0.2$ & $-14.45 \pm 0.01$ & $1.70 \pm 0.03$ & 0.07 & 0.27 & $3.1 \times 10^4$\\
CL3 ............ & 17 12 06.48 & $-$39 31 14.7 & \phantom{0}$9.6 \pm 0.4$ & $-12.50 \pm 0.02$ & $0.95 \pm 0.04$ & 0.06 & 0.08 & $1.6 \times 10^4$\\
CL4 ............ & 17 11 57.33 & $-$39 32 08.1 & \phantom{0}$4.4 \pm 0.2$ & $-12.32 \pm 0.05$ & $2.79 \pm 0.12$ & 0.10 & 0.33 & $1.2 \times 10^4$\\
CL5 ............ & 17 11 53.17 & $-$39 31 21.3 & $13.5 \pm 0.9$ & $-11.99 \pm 0.05$ & $1.38 \pm 0.11$ & 0.10 & 0.47 & $2.0 \times 10^4$\\
CL6 ............ & 17 11 52.72 & $-$39 33 35.2 & \phantom{0}$6.0 \pm 0.6$ & $-11.49 \pm 0.06$ & $1.09 \pm 0.14$ & 0.07 & 0.08 & $1.0 \times 10^4$\\
\enddata
\tablecomments{Col. (1): Cloudlet name. Cols. (2--9): Observed properties of cloudlets obtained by single Gaussian fitting with $^{12}$CO($J$=1--0) emission line. Cols. (2)--(3): Position of cloudlets. Col. (4): Maximum brightness temperature. Col. (5): Central velocity. Col. (6): FWHM linewidth. Col. (7): Effective diameter of cloudlets defined as $(S / \pi)^{0.5} \times 2$, where $S$ is the surface area of cloudlets surrounded by a contour of the half level of maximum integrated intensity. Col. (8): Mass of cloudlets defined as $m_\mathrm{p} \mu \Omega D^2 \sum_{i} [N_\mathrm{i}(\mathrm{H_\mathrm{2}})]$, where $m_\mathrm{p}$ is the atomic hydrogen mass, $\mu$ is the mean molecular weight, $D$ is the distance to the SNR, $\Omega$ is the solid angle in a spatial pixel, and $N$($\mathrm{H_2}$) is molecular hydrogen column density for each pixel. We used $\mu = 2.8$ and an equation of $N(\mathrm{H_2}) / W(\mathrm{CO}) = 2.0 \times 10^{20}$~(K~km~s$^{-1}$)$^{-1}$~cm$^{-2}$, where $W$(CO) is the CO integrated intensity \citep{1993ApJ...416..587B}. Col. (9): Number density of molecular hydrogen $n(\mathrm{H_2})$.}
\label{tab:mc}
\end{deluxetable*}

Figure \ref{fig1}b shows the $^{12}$CO($J$=1--0) integrated intensity map {obtained using ALMA (see Appendix \ref{subsec:alma} for detailed information of observations and data reduction).} The integration range of $-16$--$-6$~km~s$^{-1}$ is typical radial velocity in the northwestern molecular clouds, which is certainly associated with RX~J1713.7$-$3946 \citep[e.g.,][]{2003PASJ...55L..61F,2012ApJ...746...82F,2005ApJ...631..947M,2010ApJ...724...59S,2013ApJ...778...59S,2015ApJ...799..175S,2012MNRAS.422.2230M}. We newly identified three molecular complexes with a size of $\sim$1~pc toward the southeastern half of the ALMA observed area, in addition to the previously known molecular cloud peak D. The X-ray bright filaments {as shown by contours} are nicely along not only with the three molecular complexes {(hereafter RXJ1713NW-East)}, but also with a part of the molecular cloud peak D {(hereafter RXJ1713NW-West)}. It is noteworthy that intercloud diffuse regions show significantly low intensity (purple colored areas), whereas there are several tiny CO clumps with a size on the order of 0.01~pc. We hereafter refer to the tiny CO clumps as ``molecular cloudlets.''

Figures \ref{fig2}a--\ref{fig2}c show the velocity channel maps of ALMA CO toward {RXJ1713NW-East}. We confirmed that one of the newly identified molecular complexes is consisting of dozens of molecular cloudlets (e.g., CL1--6) and filaments (e.g., Figure \ref{fig2}b, $\sim$0.06~pc width). Almost all the molecular cloudlets and filaments are spatially associated with the X-ray filaments, {but the X-ray minor peaks and X-ray hotspots identified by \cite{2020ApJ...899..102H} are located in the intercloud regions.} The typical spatial separations between X-ray {minor peaks or} hotspots and nearest CO intensity peaks are $\sim$0.05--0.15~pc. Figure \ref{fig2} bottom panels show CO line profiles of the typical molecular cloudlets CL1--6 {which are located near the X-ray filaments and hotspots without contamination}. All the sampled molecular cloudlets are significantly detected, and show the narrow velocity width ($< 2$~km~s$^{-1}$) except for CL4. Although wing-like profiles are seen toward CL1 and CL5, these can be explained by other overlapped molecular cloudlets. The physical properties of molecular cloudlets CL1--6 are summarized in Table \ref{tab:mc}. The typical diameters and masses of molecular cloudlets are $\sim$0.06--0.10~pc and $\sim$0.1--0.5~$M_{\odot}$, respectively. The number densities of molecular cloudlets are a few 10$^4$~cm$^{-3}$, which were obtained using a CO-to-H$_2$ conversion factor (see Table \ref{tab:mc} note).

\begin{figure*}[]
\begin{center}
\includegraphics[width=\linewidth,clip]{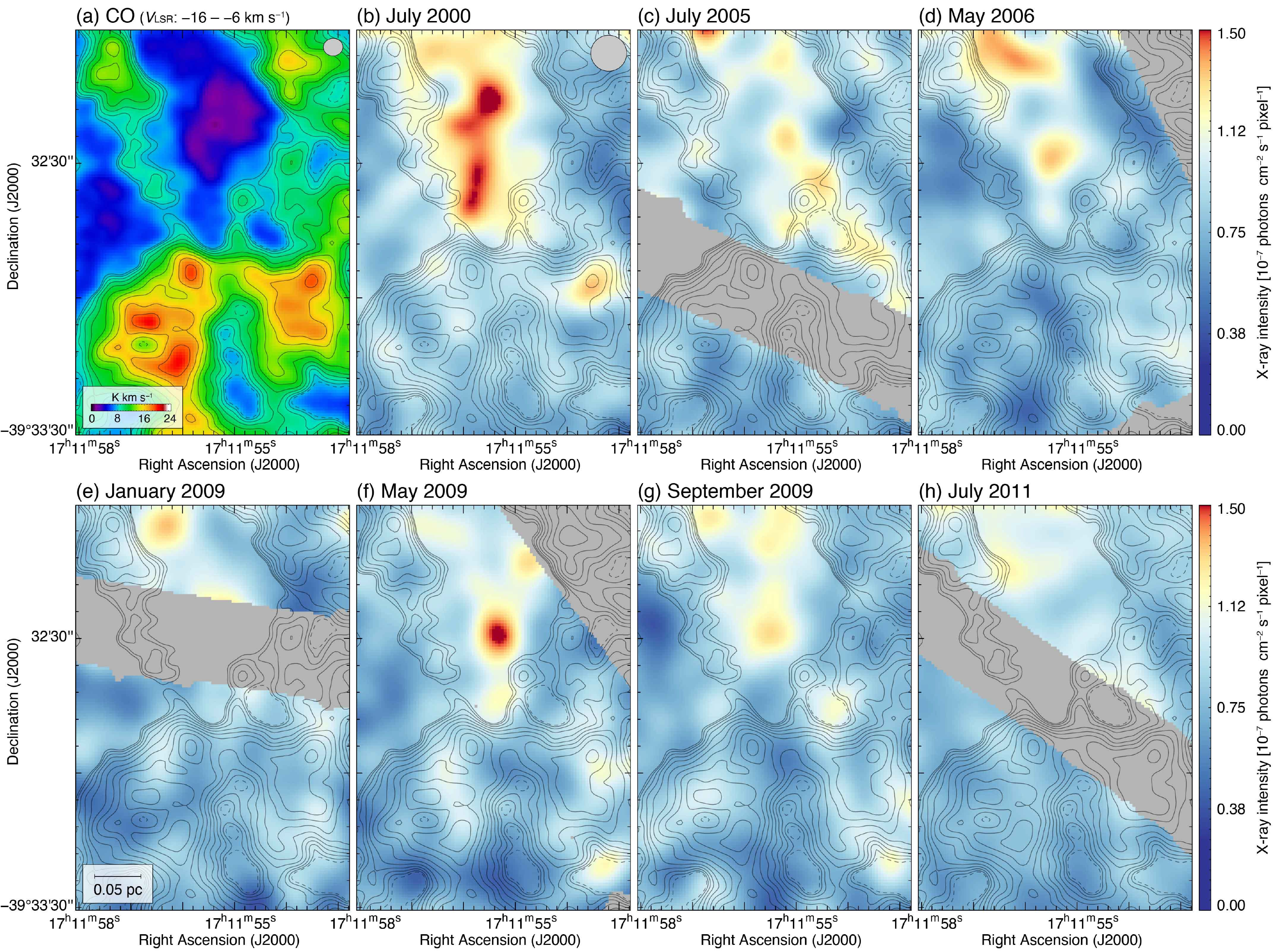}
\caption{(a) Enlarged view of ALMA CO map toward an X-ray hotspot presented by \cite{2007Natur.449..576U}. (b--h) Sequence of X-ray observations in July 2000, May 2006, May 2009, and September 2009. All X-ray images show the same intensity scales. Superposed contours indicate the CO intensity. Gray shaded areas were eliminated due to low exposure time (see the text).} 
\label{fig3}
\end{center}
\end{figure*}

Figure \ref{fig3} shows the enlarged view of the X-ray filament from which flux time variability was reported by \cite{2007Natur.449..576U}. We compared spatial distributions of CO and X-rays for each observing epoch {(see Appendix \ref{subsec:chandra} for detailes).} {T}wo X-ray hotspots in July 2000 (Figure \ref{fig3}b) and May 2009 (Figure \ref{fig3}f) are significantly detected. The former was previously reported by \cite{2007Natur.449..576U}, whereas the latter is newly identified ($\sim$4$~\sigma$ above the surrounding level). The latter hotspot was excited within three years and disappeared within four months. It should be also noted that these hotspots are located on the intercloud or low-density region, but not in the direction of dense cloudlets.

\section{Discussion}\label{sec:discussion}
\subsection{Origin of highly inhomogeneous density distribution}\label{subsec:windbubble}
We spatially resolved shocked molecular cloudlets and filaments toward the northwestern shell of RX~J1713.7$-$3946. These clumpy structures at 0.01~pc scales coexist with the low-density intercloud medium. We argue that the highly inhomogeneous gas environment provides conclusive evidence for a wind-bubble scenario proposed by \cite{2012ApJ...746...82F} and \cite{2012ApJ...744...71I}. Before the supernova explosion, the high-mass progenitor of RX~J1713.7$-$3946 ejected its outer hydrogen layer as stellar winds over a time scale of several $10^6$ years. The pre-existent intercloud diffuse gas was completely swept up and a low-density wind-bubble with a density of $\sim$0.01--0.1~cm$^{-3}$ was created \citep[e.g.,][]{1977ApJ...218..377W}. On the other hand, molecular cloudlets and filaments can survive wind erosion due to their high density $\sim$10$^4$~cm$^{-3}$. {Moreover, according to synthetic observations of MHD simulation for colliding H{\sc i} clouds, $\sim$0.1 pc clouds could be newly formed by stellar wind compression \citep[e.g.,][]{2018ApJ...860...33F,2018arXiv181102224T}.} After the passage of the supernova shock wave, dense cloudlets and filaments will not be deformed or evaporated owing to short interacting time. In fact, numerical simulations show that a molecular cloud with the size 0.2~pc and density $10^3$~cm$^{-3}$ can survive shock erosion at least 300~years after the passage of shocks \citep[][]{2019MNRAS.487.3199C}. The CO line emission without line-broadening or wing-like profiles also supports this idea.

Note that such inhomogeneous gas density distribution is also expected in other core-collapse SNRs. Further ALMA observations with high special resolution on the order of 0.01~pc are needed for complete understanding the interstellar environment surrounding the SNRs.  

\subsection{Magnetic field amplification via shock-cloudlet interactions}\label{subsec:bamplification}
\cite{2007Natur.449..576U} discovered X-ray hotspots in the northwestern shell of RX~J1713.7$-$3946, which show year-scale time variability of X-ray flux with the typical spatial scale of $\sim$0.05~pc. Considering the acceleration and radiative cooling time of cosmic-ray electrons, the authors concluded that the time variability was caused by amplified magnetic field of mG. Most recently, \cite{2020ApJ...899..102H} presented a detailed analysis of 65 X-ray hotspots; about one-third of them showed year-scale time variabilities and eight hotspots showed monthly variabilities with significance of at least 3$\sigma$. The authors proposed that the time variabilities are caused by dense cloud cores with a density of $10^5$--$10^7$~cm$^{-3}$. In the present section, we argue that the observed time variabilities can be understood by the magnetic field amplification through interactions between shockwaves and cloudlets with a density of $\sim$$10^4$~cm$^{-3}$.

We estimate the magnetic field strength toward the newly identified X-ray hotspot, following the method by \cite{2007Natur.449..576U} and \cite{2020ApJ...899..102H}. {The radiative cooling time of electrons $T_\mathrm{synch}$ can be written as;}
\begin{eqnarray}
T_\mathrm{synch} \sim 1.5\:\Bigl(\frac{B}{\mathrm{1~mG}}\Bigr)^{-1.5}\Bigl(\frac{\varepsilon}{\mathrm{1~keV}}\Bigr)^{-0.5}\mathrm{(year)}\;\;\;\;
\label{eq0}
\end{eqnarray}
{where $\varepsilon$ is the photon energy of synchrotron X-rays and $B$ is the magnetic field strength. Considering the short decay time of four months and $\varepsilon = 1$~keV, we can obtain $\sim$3~mG toward the X-ray hotspot. On the other hand, the acceleration time of electrons $T_\mathrm{acc}$ can be expressed in terms of gyro-factor $\eta = (B/\delta B)^2$ and shock velocity $V_\mathrm{s}$ as} 
\begin{eqnarray}
T_\mathrm{acc} \sim 1\eta\:\Bigl(\frac{B}{\mathrm{1~mG}}\Bigr)^{-1.5}\Bigl(\frac{\varepsilon}{\mathrm{1~keV}}\Bigr)^{0.5}\Bigl(\frac{V_\mathrm{s}}{3000~\mathrm{km~s^{-1}}}\Bigr)^{-2}\mathrm{(year)}.\;\;\;\;
\label{eq1}
\end{eqnarray}
{Adopting observed values of $\eta$ = 1 \citep{2019ApJ...877...96T} and $V_\mathrm{s} = 3900$~km~s$^{-1}$ \citep{2016PASJ...68..108T}, the magnetic field strength $B$ is estimated to $\sim$300~$\mu$G. Note that it is natural to have different magnetic field in Equations (\ref{eq0}) and (\ref{eq1}), because cosmic-rays do not need to be accelerated in the same location.}

We argue that the strong magnetic field in X-ray hotspots {as well as diffuse X-ray filaments} was generated by shock-cloudlet interactions. According to MHD simulations by \cite{2009ApJ...695..825I,2012ApJ...744...71I}, the shock-cloud interaction generates multiple eddies that enhance magnetic field strength up to mG on the surface of dense H{\sc i} clouds with a density of $\sim$$10^2$~cm$^{-3}$. Subsequent numerical simulation confirmed that the magnetic field is also amplified by interactions between shocks and a molecular clump with a density of $10^3$~cm$^{-3}$ \citep{2019MNRAS.487.3199C}. Note that the most enhanced magnetic field is placed $\sim$0.4~pc away from the shocked cloud by adopting the cloud size of 0.2~pc and the ISM density contrast of $\sim$$10^5$ \citep[see Figure \ref{fig3} in][]{2019MNRAS.487.3199C}. For the northwestern shell of RX~J1713.7$-$3946, the ISM density contrast is estimated to $\sim$$10^5$ considering the electron density in an intercloud region $\sim$0.10--0.13~cm$^{-3}$ \citep{2015ApJ...814...29K} and the cloudlet density $\sim$$10^4$~cm$^{-3}$. The highest magnetic field is therefore expected to be $\sim$0.12--0.20~pc away from the observed cloudlet with sizes of 0.06--0.10~pc, which is roughly consistent with observed separations between the X-ray hotspots{/filaments} and cloudlets\footnote{{Actual spatial separation between cloudlet and the highest magnetic field will be slightly modified considering the shape of shocked cloudlet and projected distance. For the case of Figure~\ref{fig3}, we assumed the cylinder shape of shocked cloud as an extension of spherical cloud assumed by \cite{2019MNRAS.487.3199C}.}}.

The shock-cloudlet interaction may also induce the spectrum modulation of synchrotron X-rays. Previous observational studies also indicate spectral flattening toward the shock interacting region \citep[e.g.,][]{2015ApJ...799..175S,2018PASJ...70...77O,2020ApJ...900L...5T}. Further spatially resolved X-ray spectroscopy based on distributions of molecular clouds reveals such spectrum modulations.

\subsection{Prospects for gamma-ray spectra toward the northwestern shell}\label{subsec:gspectrum}
It is thought that the hadronic gamma-ray shows flat $\nu F_\nu$ spectrum from the pion-creation threshold energy at $\sim$0.1~GeV to the maximum energy achieved by DSA, whereas the leptonic gamma-ray shows hard $\nu F_\nu$ spectrum instead. Since the {\it{Fermi}}-LAT detected the hard spectrum with photon index of 1.5, this was often used to reject the hadronic gamma-ray scenario from RX~J1713.7$-$3946 \citep[e.g.,][]{2011ApJ...734...28A}. On the other hand, the hadronic gamma-ray can be harder spectrum by considering the inhomogeneous gas distribution and energy dependent diffusion of cosmic-ray protons \citep[e.g.,][]{2009MNRAS.396.1629G,2010ApJ...708..965Z,2012ApJ...744...71I,2014MNRAS.445L..70G}. In this section, we argue that hadronic gamma-ray spectrum from the northwestern shell of RX~J1713.7$-$3946 will {possibly} be flatter than spatially combined spectra.

According to \cite{2012ApJ...744...71I}, the penetration depth of a relativistic particle into molecular clouds $l_\mathrm{pd}$ can be described as below;
\begin{eqnarray}
l_\mathrm{pd} =  0.1\; \eta^{0.5}\Bigl(\frac{E}{10\;\mathrm{TeV}}\Bigr)^{0.5}\Bigl(\frac{B}{100\;\mathrm{\mu G}}\Bigr)^{-0.5}
\Bigl(\frac{t}{1000\;\mathrm{yr}}\Bigr)^{0.5}\mathrm{(pc)},\;\;\;\;
\label{eq3}
\end{eqnarray}
where $E$ is the particle energy and {$t$ is the elapsed time passed since the forward shock encountered the molecular cloud.}
The magnetic field will be enhanced not only on the surface of shocked clouds via shock interaction, but also inside the clouds caused by cosmic-ray streaming what is called Bell instability \citep{2004MNRAS.353..550B}. \cite{2019ApJ...872...46I} argued that the Bell instability induced magnetic field amplification prevent further cosmic-ray penetration into shocked clouds by diminishing the diffusion coefficient for the cosmic-ray energy below $\sim$1~TeV. Note that the degree of magnetic field amplification depends on path length from the surface of shocked clouds, roughly corresponding to a radius of shocked clouds.

\begin{figure}[]
\begin{center}
\includegraphics[width=\linewidth,clip]{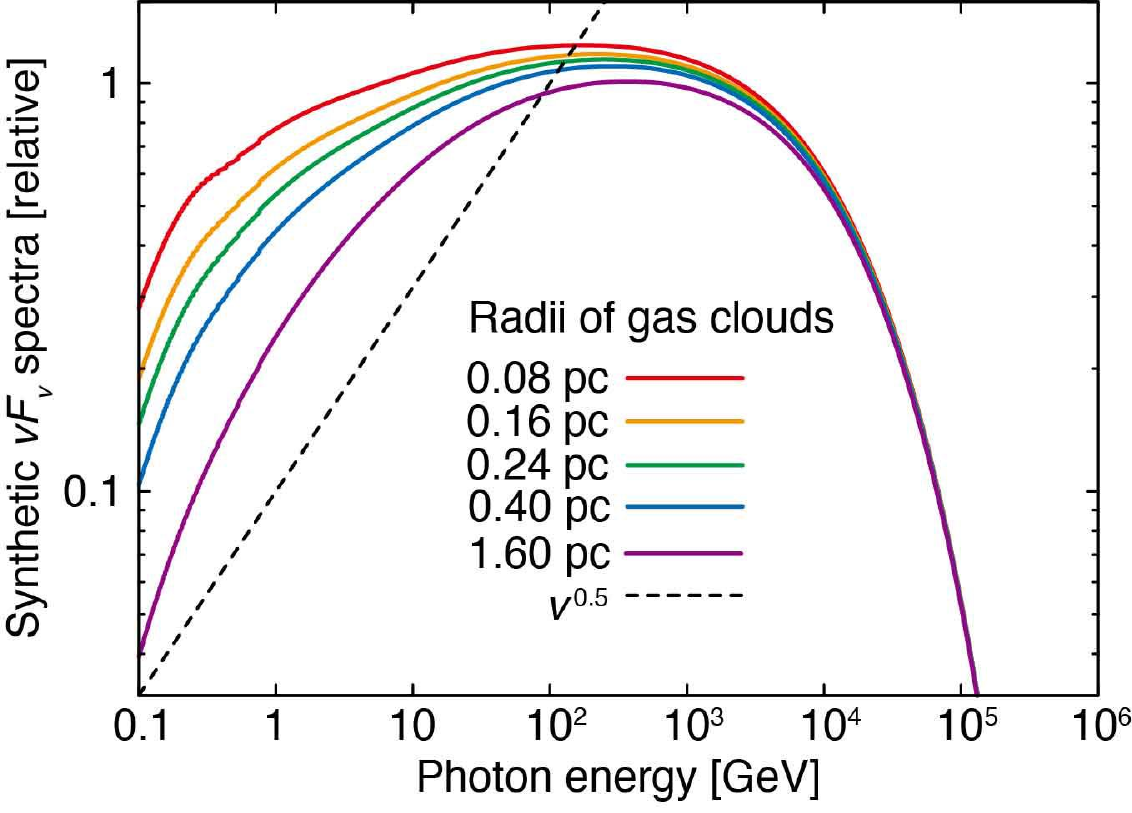}
\caption{Synthetic $\nu F_\nu$ gamma-ray spectra expected from shocked gas clouds with radii of 0.08, 0.16, 0.24, 0.40, and 1.60 pc, which were reproduced using numerical results in \cite{2019ApJ...872...46I}. The dashed line indicates $\nu F_\nu \propto \nu^{0.5}$.} 
\label{fig4}
\end{center}
\end{figure}

Figure \ref{fig4} shows $\nu F_\nu$ spectra of hadronic gamma-rays with various cloud radii. According to \cite{2005ApJ...631..947M}, typical radii of molecular clouds associated with the SNR were derived as $\sim$1--2~pc. The densest star-forming cloud peak C in the southwestern shell is a typical example with a featureless morphology and central concentration density gradient with an outer radius of $\sim$1.5~pc \citep{2010ApJ...724...59S}. Therefore, it is natural to be observed hard gamma-ray spectra ($\nu^{0.5}$ or photon index of 1.5) when we combined them for the entire remnant \citep[e.g.,][]{2011ApJ...734...28A}. On the other hand, the brightest gamma-ray peak in the northwestern shell is associated with dozens of molecular cloudlets and filaments with typical radii of 0.03--0.05~pc, suggesting that the flatter gamma-ray spectrum will be expected than the case of 0.08~pc in red curve of Figure \ref{fig4}. {Future} detailed spatial{ly} resolved spectroscopy of gamma-ray using the Cherenkov Telescope Array (CTA) has a potential to {resolve such spatial difference of gamma-ray spectral shape.}

{In case we do not detect the gamma-ray spectrum flattening in the northwestern shell of RX~J1713.7$-$3946 using CTA, it can be explained  that the magnetic field in the cloudlet surface is more efficiently enhanced than inside the cloud via the shock-cloudlet interaction. Because smaller cloud size induces shorter time scale of magnetic field amplification even if the maximum magnetic field strength does not depend on the cloud size. Alternatively, the diffusion length of cosmic-rays could be shorter than 0.1 pc if the actual elapsed time in Equation (\ref{eq3}) is significantly smaller than 1000~yrs. In fact, CO cloudlets in RXJ1713NW-East have been fully overtaken by the shocks, while the peak~D including RXJ1713NW-West is likely now interacting with the forward shock \citep[e.g.,][]{2013ApJ...778...59S}. In any case, the clumpy cloud distribution interacting with shockwaves is important in understanding the gamma-ray and X-ray spectra from shock-accelerated cosmic-rays in SNRs.}

\section{Summary}\label{sec:summary}
By using ALMA, we have spatially resolved molecular complexes consisting of the 0.01~pc scale cloudlets and filaments interacting with the northwestern shell of RX~J1713.7$-$3946, where the brightest TeV gamma-rays and synchrotron X-ray filaments/hotspots are detected. The molecular cloudlets having the typical radii of $\sim$0.03--0.05~pc and densities of $\sim$10$^4$~cm$^{-3}$ are nicely along with X-ray filaments and hotspots, indicating that magnetic field is amplified through shock-cloudlet interactions. The ISM density contrast is to be $\sim$$10^5$, consistent with a wind-bubble scenario. The small-scale structures or density fluctuations of the ISM may induce spectral modulations both the hadronic gamma-rays and synchrotron X-rays, beyond the standard DSA.

\acknowledgments
This paper makes use of the following ALMA data: ADS/JAO.ALMA\#2017.1.01406.S. {ALMA is a partnership of ESO (representing its member states), NSF (USA) and NINS (Japan), together with NRC (Canada), MOST and ASIAA (Taiwan), and KASI (Republic of Korea), in cooperation with the Republic of Chile. The Joint ALMA Observatory is operated by ESO, AUI/NRAO and NAOJ.} This study was supported by JSPS KAKENHI Grant Numbers JP18H01232 (R.Y.), JP19K14758 (H.S.) and JP19H05075 (H.S.). K. Tokuda was supported by NAOJ ALMA Scientific Research Grant Number of 2016-03B. {We appreciate the anonymous referee for useful comments and suggestions, which helped authors to improve the paper.}\\

\appendix
\vspace*{-0.5cm}
\section{Observations and Data Reductions}\label{sec:obs}
\subsection{ALMA CO}\label{subsec:alma}
Observations of $^{12}$CO($J$=1--0) line emission at 115~GHz were carried out using the ALMA Band~3 (86--116~GHz) during Cycle~5 (PI: H. Sano, proposal ID 2017.1.01406.S). We used 46 antennas of 12-m array, 12 antennas of 7-m array, and four antennas of total power (TP) array. The observed area was $11\farcm1 \times 6\farcm4$ rectangular region centered at ($\alpha_\mathrm{J2000}$, $\delta_\mathrm{J2000}$) $=$ ($17^\mathrm{h}11^\mathrm{m}48\fs0$, $-39\arcdeg30\arcmin57\farcs6$). The combined baseline length of 12-m and 7-m arrays is from 8.9 to 313.7~m, corresponding to {\it{u-v}} distances from 3.4 to 120.6~$k\lambda$. The data were processed using the Common Astronomy Software Application \citep[CASA;][]{2007ASPC..376..127M} package version 5.6.0. We utilized the ``tclean'' task with natural weighting and ``multi-scale'' deconvolution algorithm implemented in the CASA package \citep{2008ISTSP...2..793C}. We applied ``uvtaper'' during the clean processes to improve the imaging quality. We combined the cleaned interferometer data (12-m+7-m) and calibrated TP array data by using ``feather'' task. The beam size of feathered image is $4\farcs37 \times 3\farcs89$ with a position angle of $-86\fdg6$, corresponding to a spatial resolution of $\sim$0.02~pc at the distance of 1~kpc. The typical noise fluctuations of the feathered image are $\sim$0.13~K at a velocity resolution of 0.4~km s$^{-1}$.

\subsection{Chandra X-rays}\label{subsec:chandra}
We utilized archival X-ray data obtained using {\it{Chandra}}, for which the observation IDs are 736, 5560, 6370, 10090, 10091, 10092, and 12671 (PI: P. Slane for 736 and Y. Uchiyama for the others), which were published in previous papers \citep[][]{2003A&A...400..567U,2007Natur.449..576U,2003ApJ...593L..27L,2004ApJ...602..271L,2016PASJ...68..108T,2018PASJ...70...77O,2019ApJ...877...96T,2020ApJ...899..102H}. The X-ray data were taken with the Advanced CCD Imaging Spectrometer I-array on July 2000, July 2005, May 2006, January/May/September 2009, and July 2011. We used Chandra Interactive Analysis of Observations \citep[CIAO;][]{2006SPIE.6270E..1VF} software version 4.12 with CALDB 4.9.1 for data reprocessing and imaging. All the downloaded data were reprocessed using the ``chandra\_repro'' task. We then created energy-filtered, exposure-corrected images using the ``fluximage'' task for each observation, where the energy band of 1--2.5~keV. To eliminate regions with low photon statistics, we also masked areas with 60\% or less of maximum exposure time for each observation.

\end{document}